\documentclass[aps,prl,twocolumn,superscriptaddress]{revtex4}
\usepackage{bm}
\usepackage{graphicx}
\usepackage{amssymb,amsmath,amsbsy,amsgen,amsfonts}
\usepackage{dcolumn}
\usepackage{amsthm}
\usepackage{mathrsfs}
\usepackage{latexsym}
\usepackage{array}
\usepackage{amstext}
\usepackage{epsfig}
\usepackage{epstopdf} 

\usepackage{color}
\usepackage{units}
\usepackage{calrsfs}
\DeclareMathAlphabet\mathbfcal{OMS}{cmsy}{b}{n}

\newcommand{\be}{\begin{equation}}
\newcommand{\ee}{\end{equation}}
\newcommand{\ba}{\begin{array}}
\newcommand{\ea}{\end{array}}
\newcommand{\bqa}{\begin{eqnarray}}
\newcommand{\eqa}{\end{eqnarray}}

\begin{document}

\title{Spontaneous lateral atomic recoil force close to a photonic topological material}

\author{S. Ali Hassani Gangaraj}
\email{ali.gangaraj@gmail.com}
\address{School of Electrical and Computer Engineering, Cornell University, Ithaca, NY 14853, USA}

\author{George W. Hanson}
\email{george@uwm.edu}
\address{Department of Electrical Engineering, University of Wisconsin-Milwaukee, 3200 N. Cramer St., Milwaukee, Wisconsin 53211, USA}

\author{Mauro Antezza}
\email{mauro.antezza@umontpellier.fr}
\address{Laboratoire Charles Coulomb (L2C), UMR 5221 CNRS-Universit\'{e} de Montpellier, F-34095 Montpellier, France}
\address{Institut Universitaire de France, 1 rue Descartes, F-75231 Paris Cedex 05, France}

\author{ M\'ario G. Silveirinha}
\email{mario.silveirinha@co.it.pt}
\address{Laboratoire Charles Coulomb (L2C), UMR 5221 CNRS-Universit\'{e} de Montpellier, F-34095 Montpellier, France}
\address{Instituto Superior T\'{e}cnico, University of Lisbon
    and Instituto de Telecomunica\c{c}\~{o}es, Torre Norte, Av. Rovisco
    Pais 1, Lisbon 1049-001, Portugal}

\date{\today}

\begin{abstract}
We investigate the quantum recoil force acting on an excited atom
close to the surface of a nonreciprocal photonic topological
insulator (PTI). The main atomic emission channel is the
unidirectional surface-plasmon propagating at the PTI-vacuum
interface, and we show that it enables a spontaneous lateral recoil
force that scales at short distance as $1/d^4$, where $d$ is the
atom-PTI separation. Remarkably, the sign of the recoil force
is polarization and orientation-independent, and it occurs in a translation-invariant homogeneous system in thermal
equilibrium. Surprisingly, the recoil force persists for very small
values of the gyration pseudovector, which, for a biased plasma,
corresponds to very low cyclotron frequencies. The ultra-strong
recoil force is rooted on the quasi-hyperbolic dispersion of the
surface-plasmons. We consider both an initially excited atom and a continuous pump
scenario, the latter giving rise to a continuous lateral force whose
direction can be changed at will by simply varying the orientation
of the biasing magnetic field. Our predictions may be tested in
experiments with cold Rydberg atoms and superconducting qubits.
\end{abstract}

\maketitle


The force on neutral atoms or nanoparticles due to electromagnetic
fields is an important tool in atomic control and manipulation
(e.g., in optical trapping for ultracold gases
\cite{Nobel1,Nobel2,Nobel3}, optical tweezers \cite{CAT1,CAT}, etc).
Also fluctuations of the electromagnetic field, of both quantum and
thermal origin, play an important role in this context. They explain
the presence of the Casimir-Polder forces acting on atoms located
close to the surface of a body \cite{CasimirPolder,DLP}. For planar
surfaces the Casimir-Polder force is along the normal direction and
can be either attractive or repulsive
\cite{neq_Mauro,Exp_Cornell,MarioCasimir,Capasso,Boyer,Stas}. The
normal component of the force has been extensively investigated both
theoretically and experimentally (see \cite{Mauro_revsm} and
references therein).

For excited systems (e.g., an excited or driven atom) it is also
possible to have nonzero lateral forces, even when the material
surface is smooth and translation invariant. Such lateral forces
have been studied both in classical and in quantum scenarios: e.g.,
when a polarizable nanoparticle is near an anisotropic substrate
with a tilted axis \cite{I2017}, when linearly-polarized light
illuminates a chiral particle \cite{C2014}, when the emitter has
circular polarization \cite{Z2013}-\cite{Z2015}, or in moving
systems \cite{Mario_moving, MarioPRX}. Such research opens new
interesting possibilities to tune and sculpt the surface-atom
interactions.

The discovery of topological light states is perhaps the most
exciting development in photonics in recent years. Starting with the
seminal studies by Haldane and Raghu \cite{Haldane1,Haldane2}, it
has been shown that some nonreciprocal photonic platforms are
inherently topological \cite{MIT1,MIT2,MIT3}, and thereby may enable
the propagation of unidirectional topologically protected and
scattering-immune edge states. More recently, it has been shown that
the concepts of topological photonics can be extended to
electromagnetic continua, with no intrinsic periodicity \cite{Silv1,
Silv2}. In particular, continuous media with broken time-reversal
symmetry, e.g., a magnetized ferrite or a magnetized electron gas,
may be understood as topologically nontrivial \cite{Silv1, Silv2,
AliMulti, NE, SZ1, NFang}.

The research of topological effects in photonic systems has been
mostly focused in the realization of devices -- such as optical
isolators or circulators -- that exploit the intrinsic ``one-way"
property of the topological states. The consequences of topological
properties in the context of quantum optics have been little
explored, with a few exceptions. For example, in
\cite{HeatTransport} it has been shown that topological states may
enable the circulation of a heat current in closed orbits in a
microwave cavity at thermodynamic equilibrium, in
\cite{arXivBuhmann} the Casimir effect with topological materials
has been characterized, and in \cite{PRA} the effect of the
topological SPP on entanglement was studied pointing to significant
advantages as compared to reciprocal systems.

In this Letter we investigate a new mechanism that enables a
lateral recoil force due to the spontaneous emission features of
atoms close to a PTI surface. In an environment invariant along the
$j$-coordinate axis, the lateral quantum recoil force
$\mathcal{F}_{j}$ associated with a spontaneous decay process
generally vanishes since the emission occurs in random directions.
For particular polarizations of the emitter (e.g., circular
polarization), the lateral force may be nonzero due to the
interference of the dipole and its image \cite{B2015, Z2015}.
However, in systems formed by reciprocal media the sign of the
lateral force is invariably polarization dependent \cite{B2015,
Z2015}.

In contrast, here we demonstrate that the PTI's unidirectional light
states create the opportunity for unusual optical manipulations of
two-level quantum systems with a strongly asymmetric, tunable, and
\emph{polarization and orientation}-independent lateral
optical force. To our best knowledge, this is the first proposal of
a lateral recoil force on an excited atom near a laterally-invariant
substrate, with the sign of the force independent of the
polarization state and of the atom orientation.

We first consider a two-level system (called the 'atom' in the
following) in a general inhomogeneous environment. In the dipole
approximation, the optical force operator is $\hat{\mathcal{F}_i} = \hat{\boldsymbol{\mathrm{p}}} \cdot
\partial_i \hat{\boldsymbol{\mathrm{E}}} $, $i=x,y,z$ \cite{CAT}.
The hat indicates that the relevant quantity represents a quantum operator. Here, $%
\hat{\mathbf{p}} $ is the electric dipole operator for the two-level
atom. By solving the Heisenberg equations with the Markov
approximation, it is possible to show that the expectation of the
optical force is given by \cite{SM}
\begin{equation}
{{\mathcal{F}}_i}\left( t \right) = \left\langle {{{\hat
{\mathcal{F}}}_i}} \right\rangle  = {\rho _{ee}}\left( t
\right){{\mathcal{F}}_{R,i}} + \left( {1 - 2{\rho _{ee}}\left( t
\right)} \right){{\mathcal{F}}_{C,i}} \label{Force_gen}
\end{equation}%
with ${\rho _{ee}}\left( t \right)$ representing the probability of
the atom being in the excited state,
\begin{equation}
{{\mathcal{F}}_{R,i}} = 2\,{\mathop{\rm Re}\nolimits} \left\{
{{{{\bf{\tilde \gamma }}}^*} \cdot {{\left. {\left( { - i\omega
{\partial _i}{\bf{G}}\left( {{\bf{r}},{{\bf{r}}_0};\omega } \right)}
\right)} \right|}_{\scriptstyle\omega  = {\omega _0} + i{0^ +
}\hfill\atop \scriptstyle{\bf{r}} = {{\bf{r}}_0}\hfill}} \cdot
{\bf{\tilde \gamma }}} \right\} \label{Force_res}
\end{equation}
is the resonant component of the force and ${\partial _i} =
{{\bf{\hat u}}_i} \cdot {\nabla _{\bf{r}}}$. Here, $\omega _{0}$ is
the atomic transition frequency, ${\bf{\tilde \gamma }} = {\left[
{\begin{array}{*{20}{c}} {\bf{\gamma }}&0
\end{array}} \right]^T}$ is a six-vector and $\gamma$ is the dipole matrix
element. Moreover, ${{\bf{\mathcal{F}}}_{C}}= - {\nabla
_{{{\bf{r}}_0}}}{{{\mathcal{E}}_{C}}}$ is the usual Casimir-Polder
force for the ground state \cite{SM}, which vanishes for planar
surfaces along laterally-invariant directions.

The force is written in terms of the system Green function  $
{\bf{G}} \left( {{\bf{r}},{\bf{r_0}};\omega } \right)$ (a $6 \times
6$ tensor) defined in the companion article \cite{SM}, with
${{\bf{r}}_0}$ the position of the atom. Only the scattering part of
the Green function needs to be considered, because by symmetry the
self-field
does not contribute to the force \cite{SM}. For simplicity we
neglect the effect of thermal photons, which is acceptable when
$\hbar \omega _{0}$ is higher than the average thermal photon energy
$k_{B}T\ll \hbar \omega _{0}$ and when $d \ll \lambda_T$, with
$\lambda_T = hc/{k_B}T$ the thermal wavelength and $d$ the distance
between the atom and the macroscopic body. Equation
(\ref{Force_gen}) generalizes the theory of \cite{OptF1, OptF2} to
arbitrary reciprocal or nonreciprocal and possibly bianisotropic
systems.

\begin{figure}[h!]
\begin{center}
\noindent \includegraphics[width=3.5in]{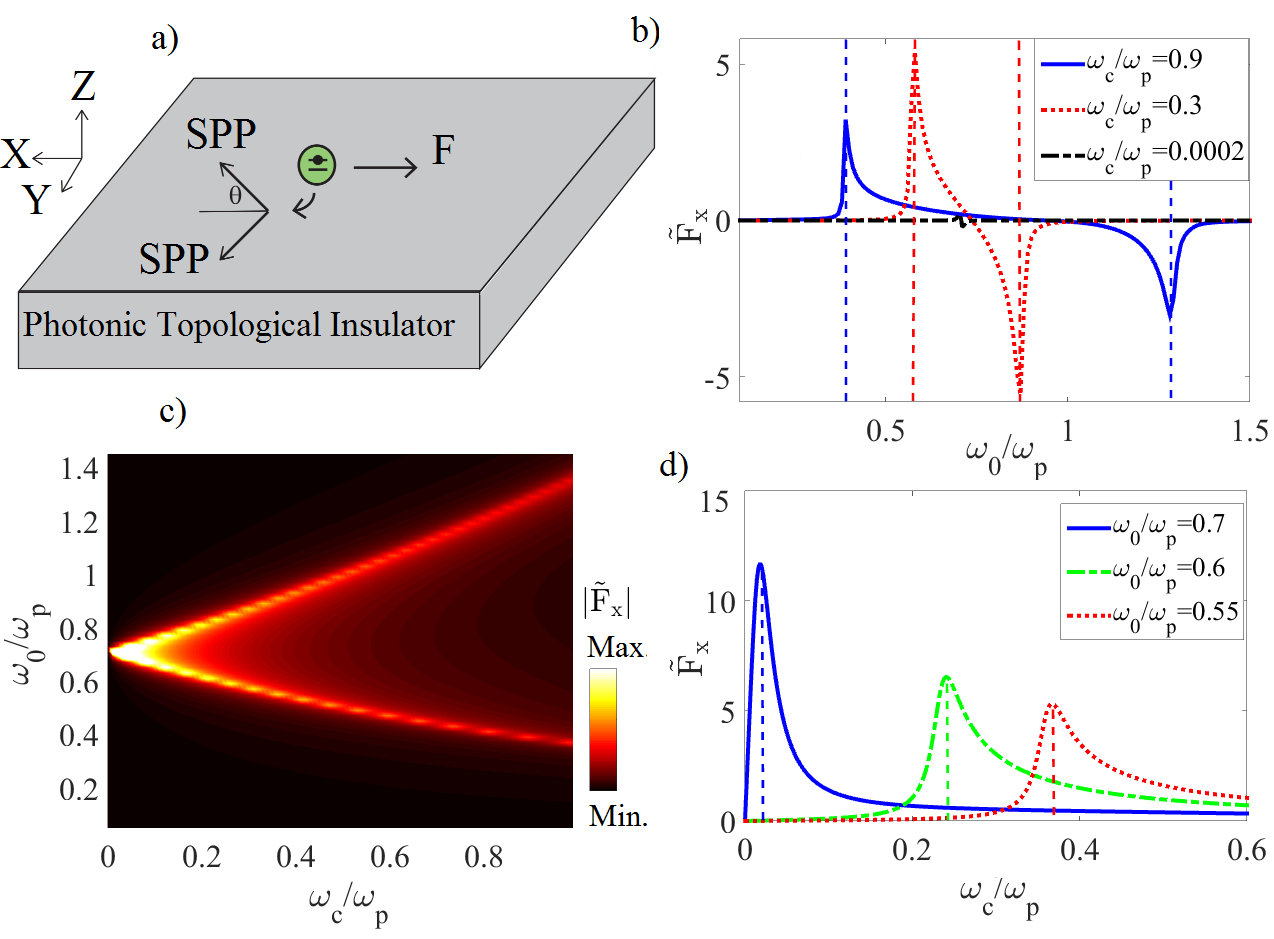}
\end{center}
\caption{{\bf (a)} The geometry under investigation. The green dot
indicates a two-level atom above an interface between a PTI
substrate and vacuum. {\bf (b)} Recoil force as a function of
frequency obtained from the exact
solution (\protect\ref{Force_res}). The vertical dashed lines indicate $\protect%
\omega_0=\protect\omega_{\pm}$ here and in panel (d). {\bf (c)}
Absolute value of the recoil force as a function of atomic
transition frequency and plasma bias. {\bf (d)} Recoil force as a
function of plasma bias for
several transition frequencies. The atom is located $d = 0.01 c / \protect%
\omega_p $ above the interface in the vacuum region.}
\label{fig1}
\end{figure}

Next, we turn to the geometry of interest, a $z$-stratified
electromagnetic environment, invariant to translations along the
$\alpha =x,y$ directions. Since the lateral force due to the
zero-point energy fluctuations vanishes, $\mathcal{F}_{C,\alpha
}=0$, the lateral force is determined uniquely by the resonant term
$\mathcal{F}_{\alpha}(t)=\rho _{ee}(t) \mathcal{F}_{R,\alpha}$. We
prove in \cite{SM} that in the limit of vanishing material loss,
$\mathcal{F}_{R,\alpha}$ can be written in terms of the
electromagnetic modes ${{{\bf{F}}_{n{\bf{k}}}}}$ of the environment
as
\begin{align} \label{Force_modal}
 \mathcal{F}_{R, \alpha }=
 \mathrm{Re}\left( i\pi \sum_{\omega _{n\mathbf{k}}>0}\omega _{n\mathbf{k}}%
\tilde{\boldsymbol{\gamma }}^{\ast }\cdot \partial _{\alpha }{\mathbf{F}}_{n%
\mathbf{k}}\otimes \mathbf{F}_{n\mathbf{k}}^{\ast }\cdot \tilde{\boldsymbol{%
\gamma }}\delta (\omega _{n\mathbf{k}}-\omega _{0})\right)
\end{align}
We use six-vector notation such that ${\mathbf{F}}_{n{\bf{k}}}
= ({\mathbf{E}}_{n{\bf{k}}} ~~
{\mathbf{H}}_{n{\bf{k}}})^{\mathrm{T}}$ represents the
electromagnetic fields. The modes are normalized as detailed in
\cite{SM}, and ${{\omega _{n{\bf{k}}}}}$ is the oscillation
frequency of the mode $n{\bf{k}}$. Note that the above result fully
takes into account the material dispersion. In the following we
consider the recoil force on an atom in vacuum close to the
interface with a topological material, e.g., a gyrotropic medium
\cite{AliMulti, HeatTransport, AliScientific, NE, SZ1}, as shown in
Fig. \ref{fig1}a.

As an example, we suppose the region $z>0$ is filled by vacuum, and
that the region $z<0$ is filled with a gyrotropic material with
permittivity
\begin{equation}
{\boldsymbol{\varepsilon }}=\varepsilon _{0}(\varepsilon _{t}{\boldsymbol{%
\mathrm{I}}}_{t}+\varepsilon _{a}\mathbf{{\hat{y}}}\mathbf{{\hat{y}}}%
+i\varepsilon _{g}\mathbf{{\hat{y}}}\times \boldsymbol{\mathrm{I}}),
\label{permittivity}
\end{equation}%
where ${\boldsymbol{\mathrm{I}}}_{t}=\boldsymbol{\mathrm{I}}-\mathbf{{%
\hat{y}}{\hat{y}}}$, with $\varepsilon_g$ being the magnitude of the
gyration pseudovector. For the gyrotropic medium we consider a
magnetized plasma (e.g., InSb \cite{Palik, GarciaVidal}). For a static bias
magnetic field along the $+y$-axis the permittivity components are
\cite{Bittencourt}
\begin{align}
\label{bp} &
{\varepsilon _t} = 1 - \frac{{\omega _p^2\left( {1 + i\Gamma /\omega } \right)}}{{{{\left( {\omega  + i\Gamma } \right)}^2} - \omega _c^2}} \nonumber \\
& {\varepsilon _a} = 1 - \frac{{\omega _p^2}}{{\omega \left( {\omega
+ i\Gamma } \right)}}, \,\,\,\, {\varepsilon _g} = \frac{1}{\omega
}\frac{{\omega _c^{}\omega _p^2}}{{\omega _c^2 - {{\left( {\omega  +
i\Gamma } \right)}^2}}}.
\end{align}
Here, $\omega _{p}$ is the plasma frequency, $\Gamma$ is the
collision rate associated with damping, $\omega
_{c}=-qB_{0}/m>0$ is the cyclotron frequency, $q=-e$ is the electron
charge, $m$ is the electron effective mass, and $B_{0}$ is the
static bias. In this work, we use a damping factor on the order of $\Gamma=0.015 \omega_p$.

In the quasi-static lossless limit, the natural modes have an
electrostatic nature $\boldsymbol{\mathrm{F}}_{n
\mathbf{k}}\approx  \left[ -\nabla \phi _{n\mathbf{k}}~~%
\boldsymbol{0}\right] ^{\mathrm{T}}$, and are surface plasmon
polariton (SPP) waves. In \cite{SM}, it is shown that the SPP
resonances have the dispersion $ \omega_{\bf{k}} =
\omega_{\theta}$ with $2\omega_{\theta} = \omega_c \mathrm{cos}
({\theta}) + \sqrt{ 2\omega_p^2 + \omega_c^2 (1 +
\mathrm{sin}^2({\theta}))}$. Here, $\theta$ represents the angle of
the SPP wave vector $\bf{k}$ (parallel to the interface) and the
$x$-axis.
In general, we have
$\omega _{-}<\omega_{\theta} <\omega _{+}$ with $\omega_{\pm} =
\omega_{\theta=0/\pi}$. The quasi-static solution describes the
resonant SPP waves with large $\bf{k}$, which are the most
influential for the light-matter interactions. Different from the
unbiased case ($\omega_c=0$), for which the SPP resonance is
direction independent, $\omega_{\theta} = \omega_p /\sqrt{2}
\equiv \omega_{\rm{spp}}$, for a biased plasma the emitted
plasmons are launched along preferred directions of space (see Fig.
\ref{fig1}a). Hence, intuitively, one may expect that the sign of
the optical force $\mathcal{F}_{x}$ is contrary to the
($x$-component) of the wave vector of the SPPs excited by the atom,
i.e. the SPPs that satisfy $\omega_{\pm \theta_0} = \omega_0$.

This heuristic argument is confirmed by a detailed calculation of
    the force (\ref{Force_modal}) that gives $\mathcal{F}_{\alpha} =
    \mathcal{F}_{0} \tilde{\mathrm{F}}_{\alpha}~ (\alpha = x,y)$, with
    $\mathcal{F}_{0}={3|\gamma |^{2}}/({16\pi d^{4}\varepsilon _{0}})$ a
    normalizing factor, and $\tilde{\mathrm{F}}_{\alpha}$ a dimensionless
    parameter given by \cite{SM}
    \begin{align} \label{QS1}
    \tilde{\mathrm{F}}_{x}=-\rho _{ee}(t)  {\left. {\frac{{{\omega_\theta }{a_\theta }\cos \theta }}{{\left| {{\partial _\theta}{\omega _\theta }} \right|}}} \right|_{\theta  = {\theta _0}}} \,
    \frac{1}{2} \left( \Gamma _{+,\theta_0 }+\Gamma _{+,-\theta_0 }\right),
    \nonumber \\
    \tilde{\mathrm{F}}_{y}=-\rho _{ee}(t) {\left. {\frac{{{\omega_\theta }{a_\theta }\sin \theta }}{{\left| {{\partial _\theta}{\omega _\theta }} \right|}}} \right|_{\theta  = {\theta _0}}} \,
    \frac{1}{2} \left( \Gamma _{+,\theta_0 }-\Gamma _{+,-\theta_0 }\right).
    \end{align}
where $a_{\theta }>0$ and $\Gamma _{+,\theta}\ge0$ are
defined in \cite{SM}. Therefore, the recoil force scales as $1/d^4$
where $d$ is the distance between the atom and the interface.
Clearly, this quasi-static result shows that the sign of the force $\mathcal{F}_{x}$ is the opposite of the sign of $\cos(\theta_{0})$, and thereby the
force is anti-parallel to the wave vector of the emitted plasmons,
\emph{independent of the dipole polarization}. In \cite{SM} it is
shown that the quasi-static force expression (\ref{QS1}) yields
results virtually identical to the exact result (\ref{Force_res}).
In this Letter, we focus on the component of the lateral force
perpendicular to the biasing magnetic field ($\mathcal{F}_{x}$). The
sign of the other lateral force component ($\mathcal{F}_{y}$) is
polarization dependent. For a vertical dipole, one has
$\mathcal{F}_{y}=0$  by symmetry.

Equation (\ref{QS1}) reveals that the lateral force is
    mainly determined by the plasmons that propagate with wave vector
    directed along either $\theta=\theta_0$ or $\theta=-\theta_0$. Thus,
    the lateral force can be written as
    $\boldsymbol{\mathcal{F}}_{L}=\boldsymbol{\mathcal{F}}_{L,\theta_0}
    + \boldsymbol{\mathcal{F}}_{L,-\theta_0}$ where
    the vector $\boldsymbol{\mathcal{F}}_{L,\pm \theta_0}$ is determined by
    (\ref{QS1}) with $\Gamma _{+, \mp \theta_0 }$ set
    identical to zero. Interestingly, the vector components
    $\boldsymbol{\mathcal{F}}_{L,\pm \theta_0}$ are anti-parallel to the
    directions $\theta=\pm \theta_0$, i.e., to the wave vector of the
    excited plasmons. This indicates that the momentum transfer is
    determined by the Minkowski momentum, rather than by the Abraham
    momentum \cite{Pfeifer, Barnett1, Barnett2, MarioMomentum}. Note
    that the Minkowski momentum is parallel to the wave vector whereas
    the Abraham momentum is parallel to the Poynting vector (or
    equivalently, to the group velocity) \cite{Barnett1, Bliokh}. As
    will be discussed later, in a gyrotropic half-space the directions of
    the plasmon wave vector and of the group velocity are generally
    different. This result suggests that light-matter
    interactions at the quantum level are determined by the canonical
    (Minkowski) momentum of light, rather than by the kinetic (Abraham)
    momentum \cite{MarioMomentum, Barnett1, Barnett2}.

Figure \ref{fig1}b shows the normalized force as a function of
$\omega _{0}$ for different bias strengths, calculated with the
exact Green function solution (\ref{Force_res}) for an atom with
 $\boldsymbol{\gamma }=\gamma
\mathbf{{\hat{z}}}$ \cite{SM}. By changing the magnetic bias
strength the recoil force can be controlled, with the force existing
primarily in the frequency interval $\omega _{-}<\omega_0 <\omega
_{+}$. Furthermore, the sign of the recoil force can be
flipped simply by flipping the bias field ($\omega_c<0$).

\begin{figure}[h!]
\begin{center}
\noindent \includegraphics[width=3.5in]{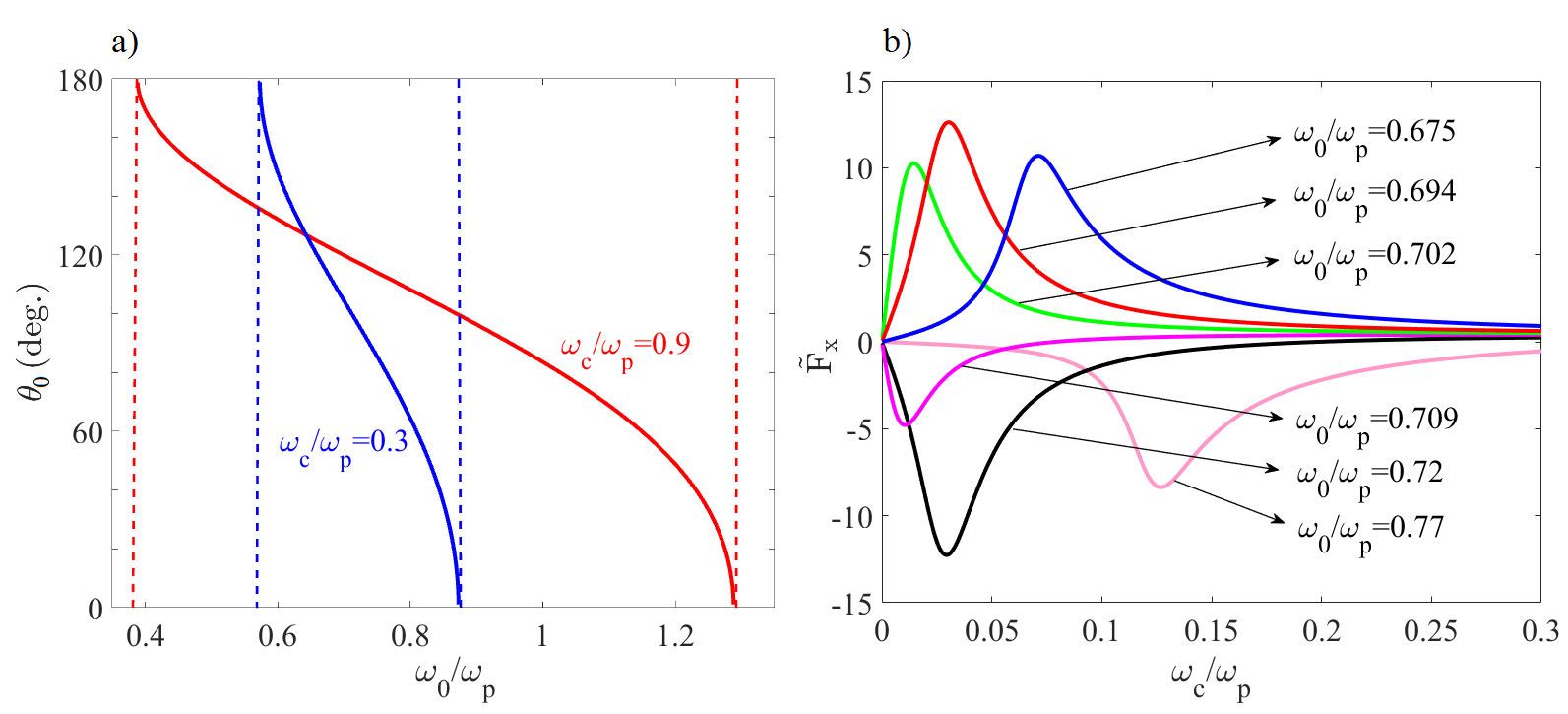}
\end{center}
\caption{ {\bf (a)} SPP wavevector angle ${\protect\theta_{0}} $ as
a
function of the atomic transition frequency for $\protect\omega_c / \protect%
\omega_p = 0.4 $. The vertical dashed lines indicate $\protect\omega_0=%
\protect\omega_{\pm}$. {\bf (b)} Recoil force as a function of
cyclotron frequency for
several transition frequencies. The atom is located $d = 0.01 c / \protect%
\omega_p $ above the interface in the vacuum region.  }
     \label{figc}
\end{figure}

The two prominent observations from Fig. \ref{fig1}b are that the
recoil force changes sign as $\omega_0$ sweeps the interval
$(\omega_-,\omega_+) $, and that the recoil force is largest for
$\omega_0 = \omega_{\pm}$, denoted by the vertical dashed lines in
the figure. Both effects can be understood from conservation of
momentum; the angle of the SPP  wavevector $\mathbf{k}$ (the main
emission channel) is shown in Fig. \ref{figc}a, where it can be seen
that for $\omega_0=\omega_-$, $\theta_{0}=180^{\circ}$
($-x$-direction), resulting in a positive lateral force along $x$,
whereas for $\omega_0=\omega_+$, $\theta_{0}=0^{\circ}$
($+x$-direction), leading to a negative lateral force. Similarly,
the recoil force will be largest for $\omega_0 = \omega_{\pm}$,
since $\tilde{\mathrm{F}}_{x}$ is maximized when the emitted SPP is
along the $\pm x$ axis.

\begin{figure}[h]
\begin{center}
\noindent \includegraphics[width=3.5in]{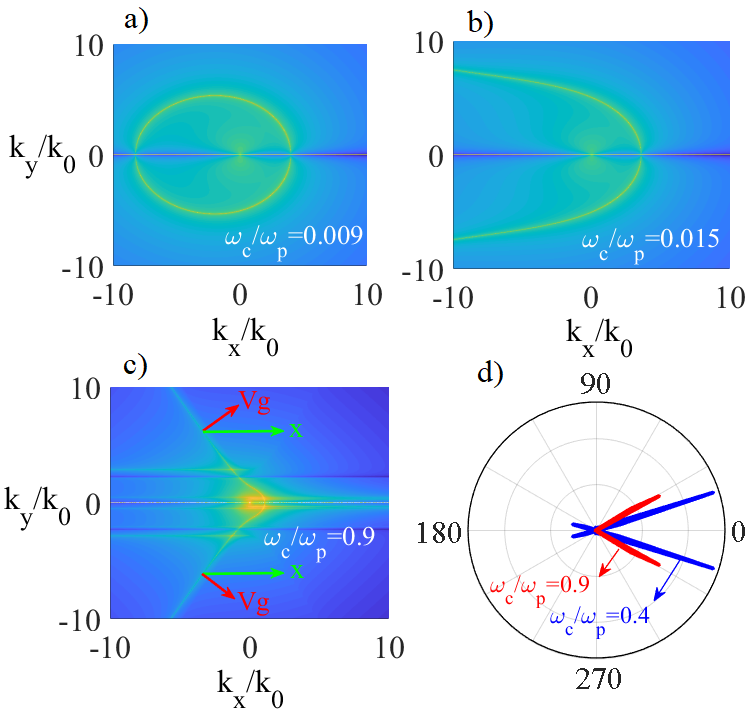}
\end{center}
\caption{{(a)-(c)} Equifrequency contours for the SPP at the
interface of a magnetized
plasma and vacuum, for different bias strengths, with $\protect\omega/\protect%
\omega _{p}=0.7$ and $k_0=\omega/c$. The red arrows in panel c)
indicate the direction of SPP energy flow. {(d)} SPP far-field
pattern (radiated by a classical dipole polarized along $z$) for two
bias values.} \label{fig5}
\end{figure}

A further aspect of the SPPs is that they form beam-like far-field
patterns due to the hyperbolic nature of the material. Figure
\ref{fig5} shows the SPP equifrequency contour (EFC) \cite{SM} for
the case of $\omega _{0}/\omega _{p}=0.7$ for various values of
$\omega_{c}/\omega _{p}$. For an anisotropic medium
the energy flow direction defined by the group velocity, $%
\boldsymbol{\nabla }_{\boldsymbol{\mathrm{k}}}\omega_{\boldsymbol{\mathrm{k}}}%
$ (orthogonal to the equifrequency contour) does not necessarily
coincide with the direction of the plasmon wave vector
$\mathbf{k}=({k_{x}},{k_{y}},0)$. For the case of $\omega _{c}/\omega _{p}=0$ (non-biased plasma), the EFC
is a circle, such that at each point $({k}_{x},{k}_{y})$ on the EFC
the flow of energy is along the direction normal to the circle, and
thereby the SPP emission is omnidirectional when $\boldsymbol{\gamma
}=\gamma \mathbf{{\hat{z}}}$.

For a very weak value of the bias (Fig. \ref{fig5}a), the circle
becomes slightly elongated along the ${k}_{x}$-axis. For a larger
bias the circle opens up (Fig. \ref{fig5}b), and as the bias is
increased further the EFC becomes hyperbolic-like, leading to a
unidirectional (nonreciprocal) SPP beam with energy flow indicated
by the red arrows in Fig. \ref{fig5}c. The SPP far-field pattern at
two values of bias is shown in Fig. \ref{fig5}d, where it can be
seen that the SPP forms two narrow beams. Although SPPs on a biased
plasma have been long-studied (see, e.g., \cite{1962}), this
narrow-beam, angle-dependent aspect of the SPP has not been
predicted previously.

Remarkably, a strong recoil force can persist even for very small
values of the gyration pseudovector (e.g., cyclotron
frequency/bias). Figure \ref{fig1}c shows the recoil force as a
function of magnetic bias and atomic transition frequency, where,
for a given cyclotron frequency $\omega_c$, the bright areas
correspond to $\omega_0 = \omega_{\pm}$. Figure \ref{fig1}d depicts
the recoil force as a function of the cyclotron frequency for
different atomic transition frequencies (i.e., three horizontal
slices from Fig. \ref{fig1}c), showing that for each $\omega_0$
there is an optimal value for the bias (vertical dashed line) that
maximizes the recoil force ($\omega_0 = \omega_{\pm}(\omega_c)$). As
$\omega_0 \rightarrow \omega_{\rm{spp}}$ an ultra-strong
recoil force is obtained for very small bias, as shown in detail in
Fig. \ref{figc}b.

\begin{figure}[h!]
\begin{center}
\noindent \includegraphics[width=3.5in]{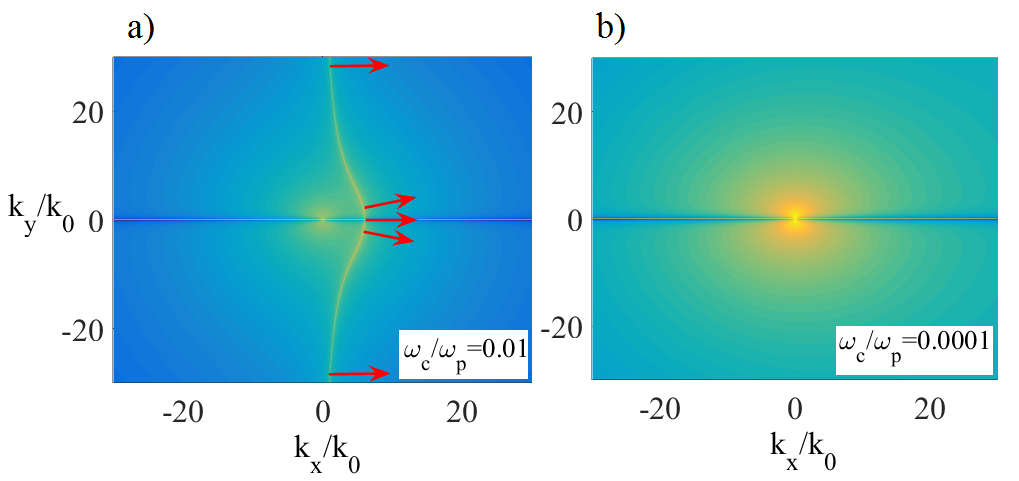}
\end{center}
\caption{SPP EFS for $\protect\omega_0 = \protect
\omega_{\rm{spp}}$ for different values of bias strength. The
red arrows indicate the direction of SPP energy flow.} \label{fig6}
\end{figure}

In order to understand the phenomenon of strong force with small
bias, it can be observed that as the bias field approaches zero, the
SPP frequency span $\omega _{-}<\omega <\omega _{+}$, becomes
narrower and narrower, as $\omega _{+}$ and $\omega _{-}$ approach
$\omega_{\rm{spp}}$ (see, e.g., Fig. \ref{figc}a), the point
at which the two bright lines in Fig. \ref{fig1}c intersect. Figure
\ref{fig6} shows the EFC of the corresponding SPP for $\omega
_{0}=\omega_{\rm{spp}}$ for small plasma bias. As can be seen in Fig. \ref{fig6}a, for small bias the EFC hyperbola branches become aligned along the 
${k}_{y}$-axis. In particular, for Fig. \ref{fig6}a there is a wide
spectral region near the center of the EFC (roughly $|{k}_{y}|/k_0
<10$), for which ${\mathrm{k}}_{x} \ne 0$, and a strong
force contribution exists. Thus, the ultra-strong force at a weak
bias is due to the high-density of states determined by the
hyperbolic EFC.

Further support for a strong recoil force at low bias when $\omega_0
\rightarrow \omega_{\rm{spp}}$ comes from the quasi-static
solution (\ref{QS1}); in the limit of no bias, it can be shown that
$ \omega_{\theta} \rightarrow \omega_{\mathrm{spp}}+
\omega_c/2 \cos \theta$ and $a_{\theta} \rightarrow 1/2$.
Then, for a $z$-polarized atom ($ \Gamma_{+,\theta} = 1 $)
(\ref{QS1}) becomes 
\begin{align}
\tilde{\mathrm{F}}_{x}  &=-\rho_{ee}(t) \frac{{{\omega
_{{\rm{spp}}}}}}{{{\omega _c}}}\frac{{{\omega _0} - {\omega
_{{\rm{spp}}}}}}{{\sqrt {{{\left( {\omega _c^{}/2} \right)}^2} -
{{\left( {{\omega _0} - {\omega _{{\rm{spp}}}}} \right)}^2}} }}.
\label{ForceWeakLat}
\end{align}
Therefore, the quasi-static force diverges when $\left| {{\omega _0}
- {\omega _{{\rm{spp}}}}} \right| = \left| {{\omega _c}} \right|/2$,
i.e., when the atomic transition frequency matches $\omega_{\pm}$.
Although material absorption ensures that the exact force is
finite, the divergence of the quasi-static result indicates a strong
force persisting as $\omega_c \rightarrow 0$. However, below a
critical bias value (Fig. \ref{fig6}b), the system tends towards
isotropic (i.e., $\varepsilon _{g}\ll \varepsilon _{t}$), the
interface does not support SPP modes, and the force becomes weak. 

In the absence of an external excitation, the spontaneous emission
depopulates the atom, leading to zero steady-state force. To achieve
a steady-state recoil force, the atom can be pumped by a laser.
Starting with the master equation for a general non-reciprocal,
lossy, and inhomogeneous environment, the atomic population is found
to be \cite{SM2}
\begin{align} \label{pump}
& \rho_{ee}(t) = \frac{2 {\tilde{\Omega}^2}}{ \lambda_1 \lambda_2 }
+ \frac{ \lambda_1 (\lambda_1 + 1/2) + 2 {\tilde{\Omega}}^2 }{
\lambda_1 ( \lambda_1 - \lambda_2 ) }
e^{\lambda_1 \tilde{t}}  \nonumber \\
& ~~~~~~~~~~ + \frac{ \lambda_2 (\lambda_2 + 1/2) + 2 {\tilde{\Omega}}^2 }{
\lambda_2 ( \lambda_2 - \lambda_1 ) } e^{\lambda_2 \tilde{t}},
\end{align}
where ${\tilde{\Omega}}=\Omega/\Gamma$, with $\Omega =
\boldsymbol{\gamma} \cdot \mathbf{E} / \hbar$ the Rabi frequency and
$\Gamma$ the decay rate, $\tilde{t}=\Gamma t$, and where $
\lambda_{1,2} = \frac{1}{2} \left(-3/2 \pm \sqrt{ 1/4 -
16{\tilde{\Omega}}^2} \right)$.
 Under steady state conditions $t \rightarrow \infty $, $%
\rho_{ee,ss} = 4 {\tilde{\Omega}}^2 / ( 1+ 8 {\tilde{\Omega}}^2 ) $.

\begin{figure}[t!]
\begin{center}
\noindent \includegraphics[width=3.0in]{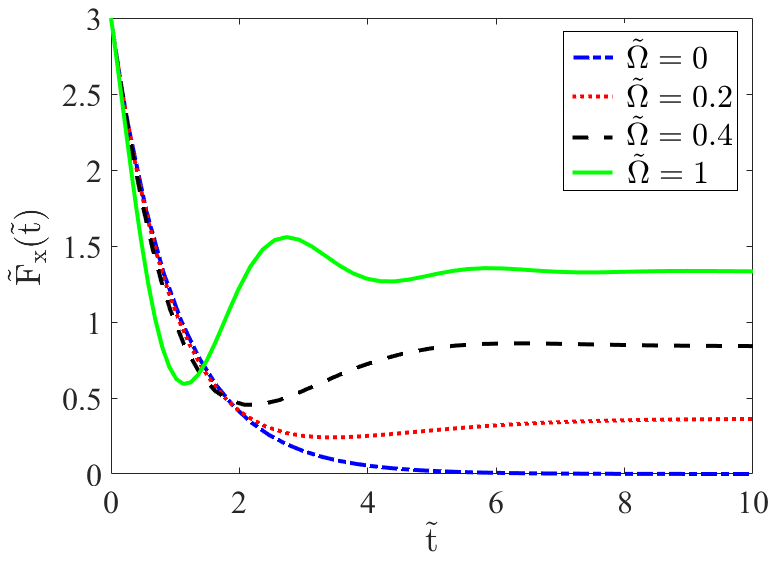}
\end{center}
\caption{Normalized recoil force as a function of time for the case of $%
\protect\omega_0 / \protect\omega_p = 0.6 $, $\protect\omega_c / \protect%
\omega_p = 0.4 $ with the atom at $d= 0.01 c / \protect\omega_p $ above the
magnetized plasma in vacuum. }
\label{F(t)}
\end{figure}

Figure \ref{F(t)} shows the normalized force $\tilde{\mathrm{F}}_x =
\mathcal{F}^\mathrm{ss}_x (t) /\mathcal{F}_0$ as a
function time for different laser intensities. As can be seen, when $\Omega
/ \Gamma = 0 $ the force decays to zero whereas for the pumped cases
non-zero steady state force exists. The laser can be applied orthogonal to the $x$-axis, so as not to influence the net lateral force.

To provide a quantitative estimate of the recoil force,
the normalization constant
$\mathcal{F}_0$ is $0.075$ pN $(\tilde{D}/{\tilde{d}}^4)$, where
$\tilde{D}$ is the dipole moment in debye and $\tilde{d}$ is the
atom-interface distance in nm. For example, for a Rydberg atom
\cite{RA} having $\gamma=7900$D located 100nm above the substrate,
$\mathcal{F}_0=0.047$pN. Alternatively, a superconducting qubit
\cite{SQ} attached to a cantilever could be used to detect the
force. One possible implementation of the plasma is InSb, where
$\omega_p \approx 31$ THz for the samples measured in \cite{Palik},
and $\omega_c$ ranges from $8-40$ THz for magnetic bias $1-5$ T,
respectively.


In summary, we derived an exact equation for the spontaneous
emission lateral recoil force on an excited atom near an interface
with a photonic topological medium. The lateral force emerges in a
homogeneous system in a thermal-equilibrium, and its sign is
independent of the atom polarization and orientation. In
order to understand the behavior of the recoil force, the SPPs
guided by the interface of the topological medium and vacuum have
been investigated. Due to the quasi-hyperbolic nature of the PTI,
the SPPs emitted by an excited atom form narrow beams, whose
direction is dependent on the atomic transition frequency. The force
is maximized at certain atomic transition frequencies, and persists
down to very small bias values due to the high-density of photonic
states. To achieve a steady-state recoil force, the quantum dynamics
of the atom has been studied by solving the relevant master equation
for a laser-pumped system.

\begin{acknowledgments}
The authors gratefully acknowledge discussions with S. Buhmann. This
work was partially funded by Funda\c{c}\~{a}o para a Ci\^{e}ncia e a
Tecnologia under project PTDC/EEI-TEL/4543/2014 and by Instituto de
Telecomunica\c{c}\~{o}es under project UID/EEA/50008/2013. M. S.
thanks the CNRS and the group Theory of Light-Matter and Quantum
Phenomena of the Laboratoire Charles Coulomb for hospitality during
his stay in Montpellier.
\end{acknowledgments}

\newpage
\cleardoublepage

\section*{Supplementary Material}

\setcounter{equation}{0}
\renewcommand{\theequation}{S\arabic{equation}}
\renewcommand\thefigure{S\arabic{figure}}
\setcounter{figure}{0}

\section*{Steady-state population for the pumped system}

Without external compensation of atomic depopulation, the atom
decays from an excited state to the ground state, $\rho_{ee}(t) =
e^{-\Gamma t} $, where, for a $z$-directed dipole,
\begin{align}  \label{coupling_terms}
& \Gamma =\frac{2 |\gamma|^2 \omega _{0}^{2}}{\varepsilon _{0}\hbar c^{2}}%
\mathrm{Im}\left(\mathcal{G}_{zz}(\mathbf{r}_{0},\mathbf{r}_{0},\omega
_{0})\right),
\end{align}
and so the steady state recoil force is zero. Here, $\mathcal{G}_{zz}$
is the $zz$-component of the standard electric Green dyadic defined as in
\cite{Hanson_1sm}. In order to have a non-zero steady state recoil
force, the atom can be pumped by an external laser. In the following
we derive the steady state recoil force under continuous pumping.

In a general non-reciprocal, lossy, and inhomogeneous environment,
for any arbitrary number of atoms, the master equation in the
interaction picture is \cite{Hanson_1sm}
\begin{align}
\frac{\partial {\rho}_s(t)}{\partial t} = -\frac{i}{\hbar} \left[{V}^{AF}, {%
\rho}_s(t)\right] + \mathcal{L} {\rho}_s(t)
\end{align}
where
\begin{align}  \label{Eq:Lindblad_bi}
\mathcal{L} {\rho}_{s}(t) & =  \nonumber \\
& \sum_{i}\frac{\Gamma _{ii}(\omega _{0}) }{2}\left( 2{\sigma}_{i} {\rho}%
_{s}(t){\sigma}_{i}^{\dagger }- {\sigma}_{i}^{\dagger }{\sigma}_{i} {\rho}%
_{s}(t)- {\rho} _{s}(t){\sigma}_{i}^{\dagger }{\sigma}_{i}\right)
\nonumber
\\
& + \sum^{i \neq j}_{i,j} \frac{\Gamma_{ij} (\omega_0) }{2} \left( \left[ {%
\sigma}_j {\rho_s}(t), {\sigma}^{\dagger}_i \right] + \left[ {\sigma}_i , {%
\rho_s}(t) {\sigma}^{\dagger}_j \right] \right)  \nonumber \\
& + \sum^{i \neq j}_{i,j} g_{ij} (\omega_0) \left( \left[ {\sigma}_j {\rho_s}%
(t), -i{\sigma}^{\dagger}_i \right] + \left[ i{\sigma}_i , {\rho_s}(t) {%
\sigma}^{\dagger}_j \right] \right)
\end{align}
is the Lindblad super-operator and the dissipative and coherent
coupling terms are, for linear polarization,
\begin{align}  \label{coupling_terms2}
& \Gamma _{ij}(\omega _{0}) =\frac{2\omega _{0}^{2}}{\varepsilon
_{0}\hbar
c^{2}}\sum_{\alpha ,\beta =x,y,z}\mathrm{\gamma}_{\alpha i}\mathrm{Im}\left(%
\mathcal{G}_{\alpha \beta }(\mathbf{r}_{i},\mathbf{r}_{j},\omega _{0})\right)%
\mathrm{\gamma}_{\beta j},  \nonumber \\
& g_{ij}(\omega _{0})=\frac{\omega _{0}^{2}}{\varepsilon _{0} \hbar
c^{2}}
\sum_{\alpha ,\beta =x,y,z} \mathrm{\gamma}_{\alpha i} \mathrm{Re}(\mathcal{G}%
_{\alpha\beta}(\mathbf{r}_{i},\mathbf{r} _{j},\omega_0 )) \mathrm{\gamma}%
_{\beta j}.
\end{align}

For a single atom, the above equation reduces to
\begin{align}  \label{Eq:Lindblad_bi1}
\mathcal{L} {\rho}_{s}(t) = \frac{\Gamma (\omega _{0}) }{2}\left( 2{\sigma} {%
\rho}_{s}(t){\sigma}^{\dagger }- {\sigma}^{\dagger }{\sigma} {\rho}_{s}(t)- {%
\rho} _{s}(t){\sigma}^{\dagger }{\sigma} \right).
\end{align}

The term describing the laser-atom interaction is
\begin{align}
{V}_{AF} & = - \hbar \left( \Omega e^{- i \Delta t} \sigma^{\dagger} +
\Omega^* e^{i \Delta t} \sigma \right),
\end{align}
where $\Omega = \boldsymbol{\gamma} \cdot \mathbf{E} / \hbar$ is the Rabi
frequency and $\Delta = \omega - \omega_0$ is the detuning parameter of the
laser with respect to the atom transition frequency. Considering $\Delta = 0
$, the dynamics of the density matrix elements in the basis $\left|e\right>$%
, $\left| g \right>$, are
\begin{align}  \label{diff_eq}
\frac{\partial \rho_{ee}(t)}{\partial t} & = -\Gamma \rho_{ee}(t) + i
\left(\Omega\rho_{ge}(t) - \Omega^* \rho_{eg}(t)\right),  \nonumber \\
\frac{\partial \rho_{eg}(t)}{\partial t} & = -\frac{\Gamma}{2} \rho_{eg}(t)
+ i \Omega \left( \rho_{gg}(t) - \rho_{ee}(t)\right),  \nonumber \\
\frac{\partial \rho_{ge}(t)}{\partial t} & = -\frac{\Gamma}{2} \rho_{ge}(t)
- i \Omega^* \left( \rho_{gg}(t) - \rho_{ee}(t)\right),  \nonumber \\
\frac{\partial \rho_{gg}(t)}{\partial t} & = \Gamma \rho_{ee}(t) + i
\left(\Omega^* \rho_{eg}(t) - \Omega \rho_{ge}(t) \right).
\end{align}

For a single atom, the density matrix $\rho_s $ is a $2\times 2$ Hermitian
conjugate matrix with diagonal elements representing the excited and ground
state probabilities satisfying $\rho_{ee}(t) + \rho_{gg}(t) = 1 $ and the
off-diagonal elements represent quantum interference satisfying $%
\rho_{eg}(t) = \rho^*_{ge}(t) $. Solving the above coupled differential
equations simultaneously for $\rho_{ee}(t) $ gives
\begin{align}
& \rho_{ee}(t) = \frac{2 \Omega^2}{ \lambda_1 \lambda_2 } + \frac{ \lambda_1
(\lambda_1 + \Gamma/2) + 2 \Omega^2 }{ \lambda_1 ( \lambda_1 - \lambda_2 ) }
e^{\lambda_1 t}  \nonumber \\
& ~~~~~~~~~~ + \frac{ \lambda_2 (\lambda_2 + \Gamma/2) + 2 \Omega^2 }{
\lambda_2 ( \lambda_2 - \lambda_1 ) } e^{\lambda_2 t},
\end{align}
where
\begin{equation}
\lambda_{1,2} = \frac{1}{2} \left(-3\Gamma/2 \pm \sqrt{ \Gamma^2/4 -
16\Omega^2} \right),
\end{equation}
It can be shown that under steady state conditions $t \rightarrow \infty $, $%
\rho_{ee,ss} = 4\Omega^2 / ( 8 \Omega^2 + \Gamma^2 ) $.

It should be mentioned that the laser not only provides an excitation to the
atom, but it also supplies momentum.  The time-averaged gradient force on the atom due to the laser is
\begin{equation}
\mathcal{F}^\mathrm{laser}= \frac{1}{2} \nabla \,\, \mathrm{Re}
\left\{
 \gamma^{\ast} \cdot \mathbf{E}  \right\}.
\end{equation}
Considering a linearly-polarized atom, and supposing, for simplicity, that the substrate acts as a perfect mirror, $E_z \propto 2 E_0 \mathrm{sin}(k_0 z)$, then the peak force reduces to
$\mathcal{F}^\mathrm{laser}_x= \mathrm{Re} \left\{ \gamma^{\ast}_z {k}_0
{E}_0 \right\} $. If we suppose the intensity of the laser $\Omega =
\boldsymbol{\gamma} \cdot \boldsymbol{\mathrm{E}} / \hbar $ is of
GHz order, and that the electric dipole moment of the atom is of
order $1$D (Debye), then ${E}_z $ is of the order $10^5$ V/m.
Considering a range of transition frequencies from far to near
infrared, ${k}_0 $ is of order $10^4-10^6 $ m$^{-1}$, and the force
applied to the atom by the laser beam is of order $10^{-17}-10^{-19}
$ N. The normalization constant $\mathcal{F}_0=3 |\gamma|^2 / 16\pi
d^4 \varepsilon_0$ is of order $10^{-12}$ N for an atom a few nm
above the interface. Therefore, it can be seen that the force due to
the laser can be very weak comparing to the atomic recoil force, and can
be ignored.


\end{document}